\documentclass[conference]{IEEEtran}
\IEEEoverridecommandlockouts
\usepackage{cite}
\usepackage[colorlinks=true, linkcolor=blue, citecolor=blue, urlcolor=blue]{hyperref}
\usepackage{amsmath,amssymb,amsfonts}
\usepackage{tikz}
\usetikzlibrary{fit,positioning,calc,backgrounds,decorations.pathreplacing,shapes.geometric, arrows.meta}
\usepackage{algorithmic}
\usepackage[ruled]{algorithm2e}
\usepackage{graphicx}
\usepackage{textcomp}
\usepackage{xcolor}
\usepackage[most]{tcolorbox}

\newtcolorbox{promptbox}{
  colback=gray!5,
  colframe=black!40,
  boxrule=0.4pt,
  arc=1pt,
  left=4pt,
  right=4pt,
  top=4pt,
  bottom=4pt
}

\def\BibTeX{{\rm B\kern-.05em{\sc i\kern-.025em b}\kern-.08em
    T\kern-.1667em\lower.7ex\hbox{E}\kern-.125emX}}
\begin{document}

\title{End-to-End Abstraction-Based Control with LLM-Enhanced NL-to-LTL Translation\\

\thanks{Accepted for publication in \emph{IEEE Control Systems Letters}, 2026.}

\thanks{R.~M.~Jungers is an FNRS Honorary Research Associate. This project has received funding from the European Research Council (ERC) under the European Union's Horizon 2020 research and innovation programme, grant agreement No.~864017 (L2C); from the Horizon Europe programme, grant agreement No.~101177842 (UNIMAAS); from the Walloon Region through the FLEXtest project; and from the ARC project SIDDARTA of the French Community of Belgium. A.~Bayat is supported by the French Community of Belgium through an FNRS/FRIA grant. N.~Ozay is supported in part by ARO grant W911NF-26-1-A127. A.~Abate acknowledges support from the ARIA projects ``SAINT'' and ``SuperMartingale Certificates'' and from the UKRI AI Hub on ``Mathematical Foundations of AI''.}

\thanks{Code and benchmark: \href{https://amiirbayat.github.io/NLTLBench.github.io/}{amiirbayat.github.io/NLTLBench.github.io/}.}
}

\author{\IEEEauthorblockN{Amir Bayat}
\IEEEauthorblockA{\textit{ICTEAM, UCLouvain} \\
Louvain-la-Neuve, Belgium \\
amir.bayat@uclouvain.be}
\and
\IEEEauthorblockN{Necmiye Ozay}
\IEEEauthorblockA{\textit{EECS, University of Michigan} \\
Ann Arbor, Michigan, USA \\
necmiye@umich.edu}
\and
\IEEEauthorblockN{Alessandro Abate}
\IEEEauthorblockA{\textit{Department of Computer Science} \\
University of Oxford, UK \\
alessandro.abate@cs.ox.ac.uk}
\and
\IEEEauthorblockN{Raphael M. Jungers}
\IEEEauthorblockA{\textit{ICTEAM, UCLouvain} \\
Louvain-la-Neuve, Belgium \\
raphael.jungers@uclouvain.be}
}


\maketitle

\begin{abstract}
Abstraction-Based Controller Design (ABCD) offers a principled framework
for the safe control of complex Cyber-Physical Systems (CPSs), but
interfacing real-world requirements with its formal synthesis machinery
remains a major bottleneck: such requirements are most naturally
expressed in Natural Language (NL), whereas ABCD requires formal
specifications such as Linear Temporal Logic (LTL). Large Language Models
(LLMs) offer a promising way to bridge this gap by translating NL
requirements into formal specifications. This paper makes three
contributions. First, we formalize an LLM-enhanced pipeline for ABCD, in
which NL requirements are translated into LTL and used within a formal
synthesis workflow. Second, we implement this pipeline in the
\texttt{Dionysos} toolbox and introduce a benchmark for evaluating
NL-to-LTL translation under both logical diversity and linguistic
variation. Third, experiments with state-of-the-art LLMs
reveal a consistent relationship between LTL complexity and translation
success: translation accuracy decreases as the target specification becomes
more complex in terms of Abstract Syntax Tree (AST) size, temporal depth, and
B\"uchi automaton size, while accounting for NL input length. Together,
these contributions provide both an evaluation framework and a practical
integration pathway for making ABCD more accessible while preserving the
rigor of formal methods.
\end{abstract}


\section{Introduction}
\label{sec:introduction}
Cyber-Physical Systems (CPSs) are increasingly deployed in safety-critical applications such as autonomous vehicles, medical devices, and aerospace systems. These systems rely on embedded control components to ensure that the
overall system satisfies desired objectives, including strict safety
constraints. At the same time, CPSs are expected to provide intuitive interfaces that enable non-expert users to specify complex control objectives and system requirements. Ideally, such requirements could be expressed directly in Natural Language (NL) and automatically translated into controllers that operate safely and correctly.

\begin{figure}[!t]
\centering
\resizebox{0.80\columnwidth}{!}{%
\begin{tikzpicture}[
  process/.style={
    draw=blue!60!black, fill=white, thick,
    rounded corners=4pt,
    minimum height=1.4cm, text width=5.1cm, inner sep=6pt,
    font=\small
  },
  arr/.style={
    -{Stealth[length=6pt]}, thick, blue!70!black
  },
  arroff/.style={
    -{Stealth[length=6pt]}, thick, violet!70!black
  },
  arrloop/.style={
    -{Stealth[length=6pt]}, thick, red!65!black
  },
  node distance=0.5cm
]

\node[process] (nl) {%
  \begin{minipage}[c]{1.6cm}\centering
    \includegraphics[height=1.0cm]{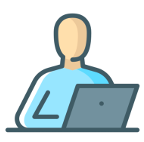}
  \end{minipage}%
  \begin{minipage}[c]{3.75cm}
    \textbf{\color{blue!70!black}NL Requirement}\\[2pt]
    {Go to blue area while avoiding brown,
    then head towards purple. Avoid obstacles all the time.}
  \end{minipage}
};

\node[process, below=0.5cm of nl] (ltl) {%
  \begin{minipage}[c]{1.4cm}\centering
    \includegraphics[height=1.5cm]{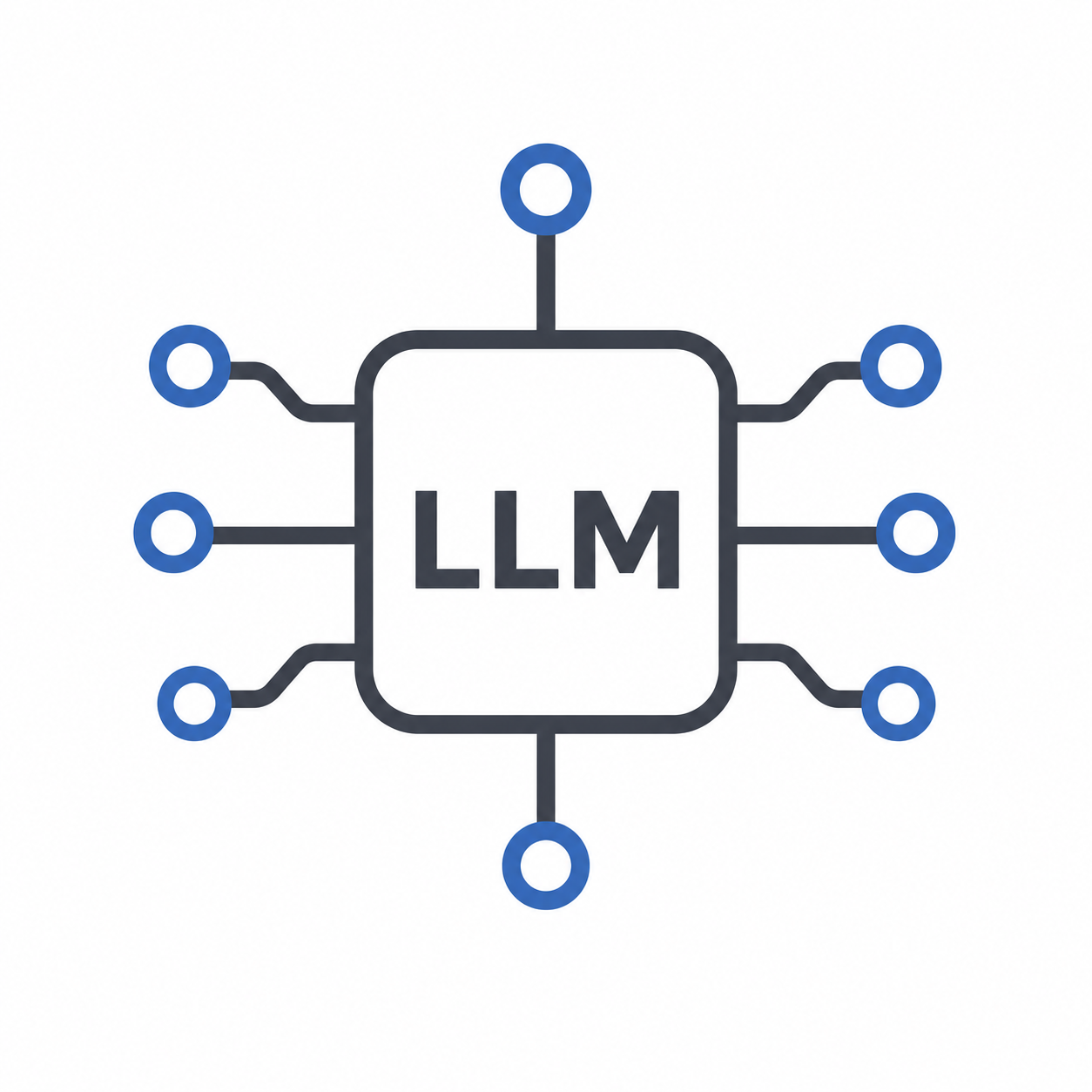}
  \end{minipage}%
  \begin{minipage}[c]{3.8cm}
    \textbf{\color{blue!70!black}NL-to-LTL Translation}\\[2pt]
    {\small
    $\bigl(\neg \textit{brown}\ \mathcal{U}\,
    (\textit{blue}\land$\\[-1pt]
    $\bigcirc\Diamond\textit{purple})\bigr)
    \land\,\Box\neg \textit{obstacle}$
    }
  \end{minipage}
};

\node[process, below=0.5cm of ltl] (explain) {%
  \begin{minipage}[c]{1.6cm}\centering
    \includegraphics[height=1.5cm]{Figs/LLM.png}
  \end{minipage}%
  \begin{minipage}[c]{3.8cm}
    \textbf{\color{blue!70!black}Explain and Validate}\\[2pt]
    {\color{blue!90!black} Back-translate the LTL formula, generate an LLM explanation, and obtain \\user validation.}
  \end{minipage}
};

\node[process, below=0.5cm of explain,
      align=center, text width=5.1cm] (buchi) {%
  \textbf{\color{blue!70!black}Construct B\"uchi Automaton with \texttt{Spot}
 and Compute the Synchronous Product}
};

\node[process, below=0.8cm of buchi,
      minimum height=1.8cm, text width=5.1cm] (ctrl) {%
  \begin{minipage}[c]{1.8cm}\centering
    \includegraphics[height=1.8cm]{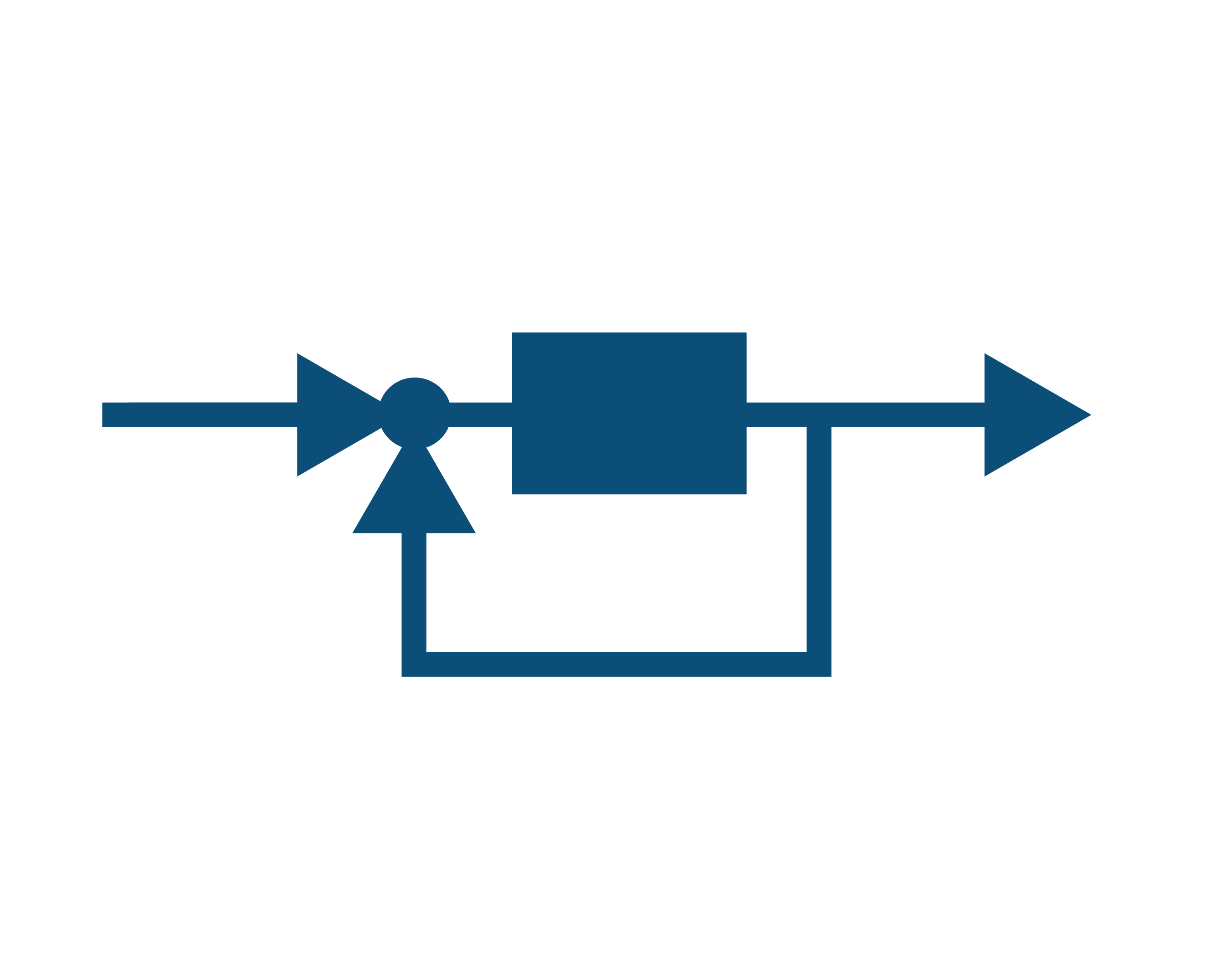}
  \end{minipage}%
  \begin{minipage}[c]{3.7cm}
    \hspace{0.6cm}
    \textbf{\color{blue!70!black}ABCD Controller}
  \end{minipage}
};

\node[draw=blue!40!black, rounded corners=5pt, thick,
      fill=white, align=center,
      at={(4.8,-0.25)},
      minimum width=2.15cm,
      font=\small] (offtitle) {%
  \textbf{Abstraction}\\[-1pt]
  \textbf{(Offline)}
};

\node[align=center, below=0.55cm of offtitle, font=\small] (cps) {%
  \includegraphics[height=2.5cm]{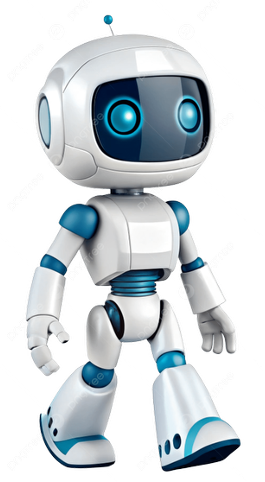}\\[-1pt]
  \textbf{Concrete System}
};

\node[align=center, below=1.47cm of cps, font=\small] (abssys) {%
  \includegraphics[height=2.5cm]{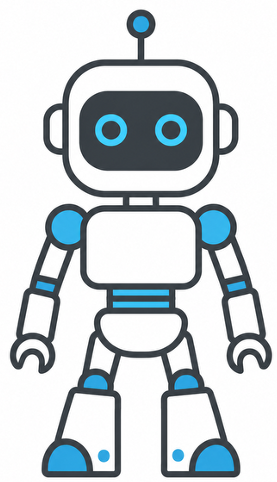}\\[-1pt]
  \textbf{Abstract System}
};

\begin{pgfonlayer}{background}
  \node[draw=blue!40!black, rounded corners=8pt, thick,
        fill=white, inner sep=5pt,
        fit=(offtitle)(cps)(abssys)] (offbox) {};
\end{pgfonlayer}

\draw[arr] (nl) -- (ltl);
\draw[arr] (ltl) -- (explain);
\draw[arr] (explain) -- node[right, font=\small\bfseries,
                             text=blue!70!black]{Valid} (buchi);
\draw[arr] (buchi) -- (ctrl);

\draw[arrloop] (explain.west)
    -- ++(-0.5,0) node[midway, above, xshift=3mm, yshift=11mm, font=\small\bfseries,
         text=red!70!black]{Invalid}
    |- (nl.west);

\draw[arrloop] (ctrl.west)
    -- ++(-0.6,0) node[midway, above,  xshift=11mm, yshift=12mm, font=\small\bfseries,
         text=red!70!black]{New requirement?}
    |- (nl.west);

\draw[arroff] (cps) -- (abssys);

\draw[arroff] (abssys.west)
    -- ++(-0.47,0)
    |- (buchi.east);

\end{tikzpicture}%
}
\caption{Overview of the proposed framework in Section~\ref{sec:ABCD_NL}. The
offline phase precomputes the abstract system from the concrete system;
the remaining steps translate an NL requirement into LTL, validate the
generated specification with the user, construct the corresponding
B\"uchi automaton, and synthesize an ABCD controller.}
\label{fig:framework}
\end{figure}

To address the complexity of CPSs while providing formal guarantees on
correctness, Abstraction-Based Controller Design
(ABCD)~\cite{tabuada2009verification}, also known as symbolic control,
provides a framework in which a system, potentially with continuous or
infinite state and input spaces, is discretized into a finite-state
symbolic abstraction according to a suitable relation. Controller synthesis is then performed on the abstract system, after which the synthesized controller is concretized to control the original system. This framework enables the design of controllers that satisfy formal specifications by construction. The abstraction step can be implemented using different strategies, depending on the structure of the system and the desired scalability. Several toolboxes have been developed to support symbolic control
synthesis~\cite{filippidis2016control,rungger2016scots,calbert2024dionysos}.
We use \texttt{Dionysos}~\cite{calbert2024dionysos}, a modular Julia-based platform supporting
multiple abstraction strategies.

Despite these advantages, specifications in symbolic control and formal methods must be expressed using formal languages, and extracting a formal representation from an NL specification requires domain expertise~\cite{gavran2020interactive}. Temporal Logics (TL), such as Linear Temporal Logic (LTL), provide precise and unambiguous representations of temporal behaviors that can be algorithmically incorporated into verification and controller synthesis frameworks~\cite{pnueli1977temporal,baier2008principles}. Since such formal specifications are difficult for non-expert users to write directly, bridging NL specifications with TL has long been an important research direction in formal methods and robotics~\cite{brunello2019synthesis}. Early translation methods relied on grammatical rules and structured language templates~\cite{kress2008translating,finucane2010ltlmop,lignos2015provably}. Later, learning-based approaches, including sequence-to-sequence models, were proposed to address the rigidity of rule-based methods and explored NL-to-TL translation~\cite{gopalan2018sequence,he2022deepstl}; however, these methods often remain sensitive to linguistic diversity and variations in phrasing.

More recently, LLMs have attracted significant attention for NL-to-TL translation. Some approaches fine-tune models such as T5 and BART~\cite{chen2023nl2tl,pan2023data}, while others rely on prompting pipelines with predefined LTL templates~\cite{fuggitti2023nl2ltl}, syntactic error correction~\cite{chen2024autotamp}, or iterative user feedback~\cite{cosler2023nl2spec}. A conformal prediction framework with statistical guarantees and human
assistance was further introduced in~\cite{sundarsingh2025conformalnl2ltl}. These methods have shown encouraging progress on NL-to-LTL translation.

Despite this progress, the field lacks a standardized evaluation
framework, and most existing works rely on bespoke datasets tailored to
specific domains, such as navigation, robot instructions, or cooking
tasks~\cite{squire2015grounding,gopalan2018sequence,mavrogiannis2024cook2ltl}. More recent
benchmark-oriented work, such as VLTL~\cite{english2025verifiable},
evaluates LLM performance in abstract NL-to-LTL translation using 43 core
LTL templates, obtained by instantiating abstract propositions and
paraphrasing the corresponding NL descriptions. Although VLTL reports
abstract NL-to-LTL translation accuracies around 95\%--100\%, its limited
template diversity provides little insight into how performance changes
with the structural complexity of the target LTL formula. This leaves open a central question for the use of LLMs in formal
specification: whether high translation accuracy persists for more
complex formulas, motivating a benchmark with broader logical coverage
and systematically varied formula complexity.

In this paper, building on our initial work~\cite{bayat2025llm}, we
leverage LLMs for NL-to-LTL translation and integrate this capability
into a seamless pipeline for ABCD from NL requirements. We implement this
pipeline in \texttt{Julia} and build on the ABCD platform
\texttt{Dionysos}~\cite{calbert2024dionysos}. As shown in
Fig.~\ref{fig:framework}, the offline phase constructs the abstract
system. The remaining steps can be executed online: NL requirements are
translated into LTL formulas, and the generated formal specification is
submitted for user validation. The pipeline then uses the \texttt{Spot}
toolbox~\cite{duret2022spot} to construct the corresponding B\"uchi
automaton and incorporates the resulting specification into the symbolic
control problem using classical formal-methods techniques.

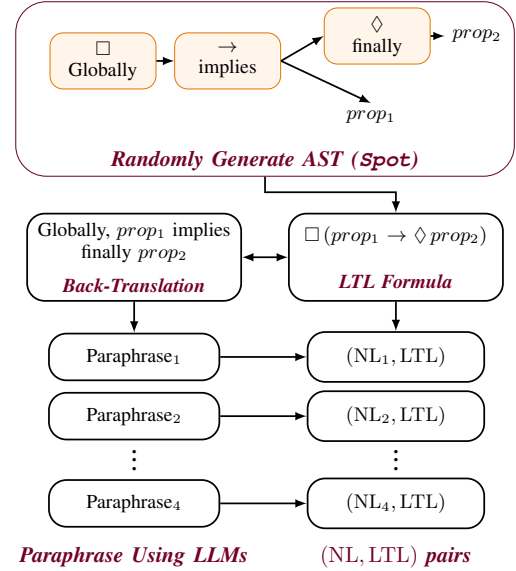
\begin{figure}[!t]
\centering
\resizebox{0.70\columnwidth}{!}{%
\begin{tikzpicture}[
    every node/.style={rounded corners=6pt, align=center, font=\small},
    operator/.style={
        draw=orange!90!black,
        fill=orange!8,
        text=black,
        minimum width=1.7cm,
        minimum height=0.9cm
    },
    leaf/.style={draw=none, fill=none, font=\small\itshape},
    box/.style={
        draw=black,
        fill=white,
        thick,
        rounded corners=8pt,
        minimum height=1.0cm,
        align=center
    },
    arrow/.style={-latex, thick, draw=black},
    biarrow/.style={latex-latex, thick, draw=black},
    labeltext/.style={
        draw=none,
        fill=none,
        font=\bfseries\itshape\color{purple!60!black}
    }
]

\node[
    draw=purple!50!black,
    fill=none,
    rounded corners=14pt,
    minimum width=8.0cm,
    minimum height=2.8cm
] (astbox) at (0,0) {};

\node[operator] (globally) at (-2.6,0.45) {$\Box$\\Globally};
\node[operator] (implies)  at (-0.6,0.45) {$\rightarrow$\\implies};
\node[operator] (finally)  at (1.8,0.85) {$\Diamond$\\eventually};

\node[leaf] (p1) at (1.7,-0.45) {$prop_1$};
\node[leaf] (p2) at (3.4,0.85) {$prop_2$};

\draw[arrow] (globally.east) -- (implies.west);
\draw[arrow] (implies.east) -- (finally.west);
\draw[arrow] (finally.east) -- (p2.west);
\draw[arrow] (implies.east) -- (p1.north);

\node[labeltext] at (0,-1.15) {Randomly Generate AST (\texttt{Spot})};

\node[box, text width=3.2cm, minimum height=1.4cm] (bt) at (-2.1,-2.7) {
    Globally, $prop_1$ implies\\
    eventually $prop_2$\\[2mm]
    {\bfseries\itshape\color{purple!60!black}Back-Translation}
};

\node[box, text width=3.2cm, minimum height=1.4cm] (ltl) at (2.1,-2.7) {
    $\Box\,(prop_1 \rightarrow \Diamond\,prop_2)$\\[4mm]
    {\bfseries\itshape\color{purple!60!black}LTL Formula}
};

\draw[arrow] (astbox.south) -- ++(0,-0.25) -| (ltl.north);
\draw[biarrow] (bt.east) -- (ltl.west);

\node[box, text width=2.5cm, minimum width=2.7cm, minimum height=0.75cm] (para1) at (-2.1,-4.3) {
    Paraphrase$_1$
};

\node[box, text width=2.5cm, minimum width=2.7cm, minimum height=0.75cm] (para2) at (-2.1,-5.25) {
    Paraphrase$_2$
};

\node[draw=none, fill=none, font=\Large] at (-2.1,-5.85) {$\vdots$};

\node[box, text width=2.5cm, minimum width=2.7cm, minimum height=0.75cm] (para4) at (-2.1,-6.65) {
    Paraphrase$_4$
};

\node[labeltext] at (-2.1,-7.5) {Paraphrase Using LLMs};

\draw[arrow] (bt.south) -- (para1.north);

\node[box, text width=2.6cm, minimum width=2.8cm, minimum height=0.8cm] (pair1) at (2.1,-4.3) {
    $(\mathrm{NL}_1,\mathrm{LTL})$
};

\node[box, text width=2.6cm, minimum width=2.8cm, minimum height=0.8cm] (pair2) at (2.1,-5.25) {
    $(\mathrm{NL}_2,\mathrm{LTL})$
};

\node[draw=none, fill=none, font=\Large] at (2.1,-5.85) {$\vdots$};

\node[box, text width=2.6cm, minimum width=2.8cm, minimum height=0.8cm] (pair4) at (2.1,-6.65) {
    $(\mathrm{NL}_4,\mathrm{LTL})$
};

\node[labeltext] at (2.1,-7.5) {$(\mathrm{NL},\mathrm{LTL})$ pairs};

\draw[arrow] (ltl.south) -- (pair1.north);

\draw[arrow] (para1.east) -- (pair1.west);
\draw[arrow] (para2.east) -- (pair2.west);
\draw[arrow] (para4.east) -- (pair4.west);

\end{tikzpicture}%
}
\caption{Benchmark construction pipeline (see
Section~\ref{sec:benchmark_generation}). Candidate LTL formulas are
generated from sampled Abstract Syntax Tree (AST) structures and filtered
for satisfiability, non-triviality, and uniqueness. Retained formulas are
back-translated into verbatim NL descriptions and paraphrased using
multiple LLMs to obtain diverse NL representations.}
\label{fig:data-generation-pipeline}
\end{figure}

In addition, to address the limitations of existing NL-to-LTL
evaluations, and in particular the limited diversity of underlying LTL
templates, we introduce a benchmark with broader logical coverage that
enables a more systematic study of LLM performance on this task. As
shown in Fig.~\ref{fig:data-generation-pipeline}, we use
\texttt{Spot} to randomly construct Abstract Syntax Trees (ASTs) with a
controlled number of nodes and then generate the corresponding LTL
formulas. The formulas are then back-translated into NL using fixed grammatical
rules, yielding phrases we call \emph{verbatim translations}. Multiple LLMs then
paraphrase these verbatim translations to increase linguistic diversity.
This allows us to study how translation performance varies across
several measures of specification and input complexity, revealing a clear
degradation in accuracy as the target specifications become more complex.

\section{Preliminaries}
\label{sec:Preliminaries}
\subsection{Linear Temporal Logic and B\"uchi Automata}
\label{sec:LTL}
LTL is a modal extension of propositional logic for reasoning about properties over discrete, linear time~\cite{pnueli1977temporal}. Let $AP=\{\textit{prop}_1,\ldots,\textit{prop}_n\}$ be a finite set of atomic propositions. LTL formulas are interpreted over infinite traces $A=A_0,A_1,A_2,\ldots$, where $A_i\in 2^{AP}$ denotes the propositions holding at time step $i$~\cite{baier2008principles}. The LTL syntax is
\begin{equation*}
    \varphi ::= true \mid \textit{prop}_{i} \mid \neg\varphi 
    \mid \varphi_{1} \wedge \varphi_{2}
    \mid \bigcirc\varphi
    \mid \Diamond\varphi
    \mid \Box\varphi
    \mid \varphi_{1} \mathcal{U} \varphi_{2},
\end{equation*}
where $\bigcirc$, $\Diamond$, $\Box$, and $\mathcal{U}$ denote the
\emph{next}, \emph{eventually}, \emph{globally}, and \emph{until}
operators, respectively. 
Here, we adopt the
\emph{abstract}\footnote{Also referred to as the \emph{lifted} setting
in the literature~\cite{chen2023nl2tl,english2025verifiable}.} setting,
where domain-specific atomic propositions are replaced by abstract
placeholders $\textit{prop}_i$. This allows us to evaluate whether models
capture the logical structure of a specification independently of
domain-specific grounding.


A B\"uchi automaton is a finite-state automaton over infinite
words, defined as
$\mathcal{B}=(Q,Q_0,\Sigma,\delta,F)$, where $Q$ is a finite
set of states, $Q_0\subseteq Q$ is the set of initial states,
$\Sigma=2^{AP}$ is the input alphabet,
$\delta\subseteq Q\times\Sigma\times Q$ is the transition relation,
and $F\subseteq Q$ is the set of accepting states. The size of
$\mathcal{B}$ is measured as $|Q|+|\delta|$~\cite{baier2008principles}.
An infinite run is accepting if it visits states in $F$ infinitely often.
For every LTL formula $\varphi$, a B\"uchi automaton
$\mathcal{B}(\varphi)$ can be constructed that accepts exactly the traces
satisfying $\varphi$. We use \texttt{Spot}~\cite{duret2022spot} to
translate LTL formulas into B\"uchi automata.

\subsection{LTL Structural Measures and Evaluation Metrics}
\label{sec:Ev}
LTL formulas can be represented as ASTs, where internal nodes correspond
to logical or temporal operators and leaves correspond to atomic
propositions or $true$~\cite{baier2008principles} (see
Fig.~\ref{fig:data-generation-pipeline}, top). The AST size of a formula
is the number of nodes in its tree, and its \emph{temporal depth} is the
maximum number of nested temporal operators along any root-to-leaf path. 

Two LTL formulas may differ syntactically while expressing the same
temporal property. Given two formulas \(\varphi\) and \(\psi\),
\emph{syntactic equality} means they share the same symbolic
representation, whereas \emph{semantic equivalence}, denoted
\(\varphi \equiv \psi\), means they are satisfied by exactly the same
traces. Since the same temporal property can be expressed by multiple
syntactically distinct LTL formulas, correctness is assessed through
semantic equivalence rather than exact string matching.

Given a benchmark set \(\mathcal{D}=\{(p_i,\varphi_i)\}_{i=1}^N\),
where \(p_i\) is an NL statement, \(\varphi_i\) is the \emph{ground-truth} LTL
formula associated with \(p_i\), and \(\hat{\varphi}(p_i)\) is the predicted formula, the
success rate is defined as
\begin{equation}
\mathrm{SR}=\frac{1}{N}\sum_{i=1}^{N}
\mathbf{1}\!\left(\hat{\varphi}(p_i)\equiv \varphi_i\right),
\label{eq:success_rate}
\end{equation}
where \(\mathbf{1}(\cdot)\) is the indicator function and equivalence is
checked using \texttt{Spot}~\cite{duret2022spot}.

\subsection{Abstraction-Based Controller Design (ABCD)}
\label{sec:ABCD}

Consider a control system with state space $X$ and input space $U$. ABCD
addresses controller synthesis over continuous or potentially infinite
state spaces by constructing a discrete, finite-state symbolic abstraction
$\mathcal{S}_a=(X_a,U_a,\rightarrow_a,AP,L)$, where $X_a$ and $U_a$ are
finite sets of abstract states and inputs,
$\rightarrow_a \subseteq X_a \times U_a \times X_a$ is the transition
relation, $AP$ is the set of atomic propositions, and
$L:X_a\rightarrow 2^{AP}$ is the labeling function. The abstraction is
constructed according to a suitable correctness relation, such as a
feedback refinement relation~\cite{reissig2016feedback}.

Given an LTL specification $\varphi$, the corresponding B\"uchi automaton
$\mathcal{B}(\varphi)$ is combined with the abstraction through the
\emph{synchronous product}.
\begin{equation}
    \mathcal{S}_P = \mathcal{S}_a \times \mathcal{B}(\varphi),
    \label{eq:product}
\end{equation}
whose states are pairs $(x_a,q)\in X_a\times Q$. A transition
$(x_a,q)\xrightarrow{u_a}(x'_a,q')$ exists whenever
$x_a\xrightarrow{u_a}_a x'_a$ in $\mathcal{S}_a$ and
$q'\in\delta(q,L(x'_a))$ in $\mathcal{B}(\varphi)$.

Controller synthesis is then performed on $\mathcal{S}_P$ using a nested fixed-point algorithm for a B\"uchi
game~\cite{tabuada2009verification}, yielding a winning set from which a
symbolic controller is extracted. The current
implementation supports LTL specifications for which the B\"uchi automaton
used for synthesis is deterministic, as required by the fixed-point
synthesis algorithm.\footnote{Further details are available
in the documentation linked on page~1.} The resulting controller ensures that
the closed-loop behavior satisfies $\varphi$ whenever the initial product
state is winning. In this work, symbolic abstraction and controller
synthesis are performed using
\texttt{Dionysos}~\cite{calbert2024dionysos}.

\section{ABCD Synthesis from NL Requirements}
\label{sec:ABCD_NL}

This section presents the proposed pipeline for synthesizing symbolic
controllers from NL requirements. As illustrated in
Fig.~\ref{fig:framework}, the offline phase first precomputes the symbolic
abstraction \(\mathcal{S}_a\) of the concrete system using predefined
abstraction parameters.

When an NL requirement \(p\) is provided by the end user, an LLM
translates \(p\) into an LTL formula \(\varphi\). The generated formula
\(\varphi\) is back-translated into an NL description using the procedure
described in Section~\ref{sec:benchmark_generation}, which is designed to faithfully reflect the semantics of
the generated formula. An LLM also provides an explanation to help the user
interpret the back-translation, while validation relies solely on
confirmation of the back-translated requirement. The user is asked
to validate the back-translated phrase before controller
synthesis is performed. This human-in-the-loop step helps ensure that the
generated formula matches the intended requirement.

After validation, the \texttt{Spot}
toolbox~\cite{duret2022spot} is used to construct the
corresponding B\"uchi automaton \(\mathcal{B}(\varphi)\). The
specification is then incorporated into the symbolic control problem by
computing the synchronous product of the abstract system and
\(\mathcal{B}(\varphi)\), as described in Section~\ref{sec:ABCD},
Eq.~\eqref{eq:product}. Controller synthesis is performed on the
resulting product system \(\mathcal{S}_P\), and the corresponding
symbolic controller is concretized to control the original system. In our
implementation, these abstraction and synthesis steps are carried out
using \texttt{Dionysos}~\cite{calbert2024dionysos}.

The pipeline also supports runtime human-in-the-loop interaction. While
the system is evolving under the synthesized controller to satisfy the
current requirement, the user may interrupt the execution and provide a
new requirement. Requirements with a finite completion time may be followed by a new
requirement once they have been satisfied.

\section{Benchmarking and Experiments}


\subsection{Benchmark Generation}
\label{sec:benchmark_generation}
As illustrated in Fig.~\ref{fig:data-generation-pipeline} and summarized
in Algorithm~\ref{alg:nltlbench_generation}, we generate a benchmark for
evaluating NL-to-LTL translation beyond a limited set of hand-written
templates. The construction allows the structural complexity of the target
LTL formulas to be systematically varied. Here, a \emph{template} is an
LTL formula pattern with fixed Boolean and temporal operator structure,
from which new formulas are obtained primarily by replacing atomic
propositions. Unlike existing datasets~\cite{english2025verifiable} that
expand a small core of such templates through proposition substitution,
our construction provides broader logical coverage, enabling a systematic
study of how translation accuracy changes with formula complexity.

The generation process is driven by benchmark \emph{configurations}
$\mathcal{C}$, which allow us to systematically vary both the logical
vocabulary and the structural complexity of the generated formulas. Each
configuration $c\in\mathcal{C}$ is a tuple
$c=(AP_c,OP_c,[n_c^{\min},n_c^{\max}],M_c)$, where
$AP_c\subseteq AP$, $OP_c\subseteq OP$,
$[n_c^{\min},n_c^{\max}]$ is the target AST-size range, and $M_c$ is the
number of candidate formulas sampled using \texttt{Spot}. Here,
$AP=\{\mathit{prop}_1,\dots,\mathit{prop}_{10}\}$ is the heuristically
chosen set of available abstract atomic propositions, and
$OP=\{\neg,\wedge,\vee,\rightarrow,\leftrightarrow,\bigcirc,\Diamond,
\Box,\mathcal{U}\}$ is the set of Boolean and temporal operators.

For each configuration, candidate LTL formulas are generated
using the \texttt{randltl} function in \texttt{Spot}~\cite{duret2022spot},
with target AST sizes sampled from the specified range. Each generated
formula is normalized by renaming its atomic propositions according to
their first occurrence, starting from $\textit{prop}_1$. This
prevents formulas differing only in proposition indices from being treated
as distinct. Each normalized formula \(\varphi\) is then filtered for
satisfiability, non-triviality, and syntactic and semantic uniqueness.

For each retained formula, we also consider its \emph{Positive Normal Form}
(PNF)~\cite{baier2008principles} when this representation is
syntactically distinct from the original formula. In PNF, negations appear only in front of atomic propositions,
while implication and equivalence are eliminated through semantically
equivalent transformations. This adds semantically
equivalent but syntactically distinct formulas, yielding additional
\emph{verbatim} NL representations produced by direct translation of the
formula structure using fixed grammatical templates.

The retained formulas and their PNF variants are translated into
verbatim NL descriptions through a rule-based back-translation procedure.
Following grammar-based logic-to-language translation
methods~\cite{pan2023data}, the procedure traverses each AST and replaces
logical and temporal operators with grammatical English templates; for
example, \(\Box\varphi\) is rendered as \emph{``always \(\varphi\)
holds''}, and \(\varphi_1 \mathcal{U} \varphi_2\) as
\emph{``\(\varphi_1\) holds until \(\varphi_2\) holds''}. The resulting
verbatim phrase \(p_{\mathrm{verb}}\) preserves the formula structure.
To obtain multiple NL representations of the same requirement, each
verbatim phrase is then paraphrased using GPT-5.4-mini, Gemini-2.5-Flash,
and DeepSeek-Chat while preserving the underlying temporal semantics,
following recent NL-to-TL benchmarks~\cite{chen2024autotamp,english2025verifiable}. Using multiple LLMs rather than repeatedly paraphrasing
with one model was intended to increase linguistic diversity.

The final dataset consists of pairs $(p,\varphi)$, where $p$ is an NL
phrase and $\varphi$ is the abstract LTL formula. Multiple
paraphrases may correspond to the same formula, enabling evaluation of
robustness under paraphrase variation. To assess semantic
faithfulness, we apply a \emph{targeted human-validation procedure}. After
the LLM evaluation in Sec.~\ref{sec:STOA_LLM}, we flag paraphrases for
which none of the evaluated LLMs produces a formula semantically equivalent
to the ground truth. This criterion only identifies cases for manual
inspection and does not imply that a paraphrase is incorrect. Flagged
paraphrases are manually reviewed and modified when necessary.

\begin{algorithm}[!t]
\caption{Benchmark Generation Algorithm}
\label{alg:nltlbench_generation}
\begin{algorithmic}[1]
\STATE \textbf{Input:} Benchmark configurations $\mathcal{C}$, each specifying
         $AP_c$, $OP_c$, $[n_c^{\min}, n_c^{\max}]$, and $M_c$
\STATE \textbf{Output:} Benchmark dataset $\mathcal{D}$

\STATE $\mathcal{D} \gets \emptyset$

\FORALL{$c \in \mathcal{C}$}
    \STATE Generate $M_c$ candidate formulas $\Phi$ with \texttt{Spot} using
    $AP_c$, $OP_c$, and AST sizes sampled from
    $[n_c^{\min}, n_c^{\max}]$
    
    \FORALL{$\varphi \in \Phi$}
        \STATE $\varphi \gets \mathrm{NormalizeAP}(\varphi)$
        
        \IF{$\varphi$ is satisfiable, non-trivial, and syntactically and semantically unique w.r.t.\ formulas in $\mathcal{D}$}
            \STATE $p_{\mathrm{verb}} \gets \mathrm{BackTranslate}(\varphi)$
            \STATE $\mathcal{D} \gets \mathcal{D} \cup \{(p_{\mathrm{verb}},\varphi)\}$
            
            \IF{$\varphi'=\mathrm{PNF}(\varphi)$ and $\varphi' \neq \varphi$ syntactically}
                \STATE $p'_{\mathrm{verb}} \gets \mathrm{BackTranslate}(\varphi')$
                \STATE $\mathcal{D} \gets \mathcal{D} \cup \{(p'_{\mathrm{verb}},\varphi')\}$
            \ENDIF
        \ENDIF
    \ENDFOR
\ENDFOR

\STATE \textbf{return} $\mathcal{D}$
\end{algorithmic}
\end{algorithm}

\subsection{Benchmark Statistics}

Following the methodology in Section~\ref{sec:benchmark_generation},
we generated 897 unique, satisfiable, and non-trivial abstract LTL formulas.
We additionally include 578 syntactically distinct PNF rewritings, yielding 1475 LTL formulas in total. Each formula is
associated with four NL variants: the verbatim sentence together with
paraphrases generated using GPT-5.4-mini, Gemini-2.5-Flash, and
DeepSeek-Chat, resulting in a total of 5900
\((\mathrm{NL},\mathrm{LTL})\) pairs. Table~\ref{tab:benchmark_summary}
summarizes the statistics of the generated benchmark and compares it with
existing datasets. 

\begin{table}[!t]
\caption{Summary statistics of the generated benchmark compared with existing benchmarks VLTL~\cite{english2025verifiable} and Conf.~\cite{sundarsingh2025conformalnl2ltl}.}
\label{tab:benchmark_summary}
\begin{center}
\setlength{\tabcolsep}{3pt}
\begin{tabular}{|p{80pt}|p{59pt}|p{30pt}|p{30pt}|}
\hline
\textbf{Statistic} &
\textbf{Our Benchmark} &
\textbf{VLTL} &
\textbf{Conf.} \\
\hline
Total LTL formulas &
1475 &
43 &
65 \\
Unique base formulas &
897 &
42 &
61 \\
Max.\ AST size &
80 &
33 &
17 \\
Max.\ \#$AP$s &
10 &
3 &
6 \\
Max.\ temporal depth &
9 &
6 &
4 \\
\hline
\end{tabular}
\end{center}
\end{table}

\subsection{LLM Performance on NL-to-LTL Translation}
\label{sec:STOA_LLM}
We evaluate several state-of-the-art LLMs on the generated benchmark
under both zero-shot and few-shot prompting, i.e., with only an
instruction prompt versus with a small number of example NL--LTL pairs
included in the prompt. We use Eq.~\eqref{eq:success_rate} as the
performance metric, counting a translation as correct when the generated
LTL formula is semantically equivalent to the ground-truth formula.

Table~\ref{tab:sota_results} reports the success rates of the evaluated
models. Claude Opus 4.7 performs best in both settings, achieving
67.69\% and 68.98\%, respectively. The targeted
human-validation procedure described in
Section~\ref{sec:benchmark_generation} rewrote 29.3\% of the paraphrases
and increased average success rate by 9.28 and 10.65 percentage points
in the two settings, respectively. 

\begin{table}[!t]
\caption{Success rates of the evaluated LLMs on our benchmark and on
the VLTL~\cite{english2025verifiable} requirements.}
\label{tab:sota_results}
\begin{center}
\setlength{\tabcolsep}{2.5pt}
\begin{tabular}{|p{65pt}|p{35pt}|p{33pt}|p{35pt}|p{33pt}|}
\hline
&
\multicolumn{2}{c|}{\textbf{Our Benchmark}} &
\multicolumn{2}{c|}{\textbf{VLTL}} \\
\cline{2-5}
\textbf{Model} &
\textbf{Zero-Shot} &
\textbf{Few-Shot} &
\textbf{\textbf{Zero-Shot}} &
\textbf{\textbf{Few-Shot}} \\
\hline
Claude Opus 4.7 & \textbf{67.69\%} & \textbf{68.98\%} & 69.05\% & \textbf{76.19\%} \\
GPT-5.5 & 62.78\% & 66.75\% & \textbf{73.81\%} & 73.81\% \\
GPT-5.4 & 59.03\% & 59.44\% & 71.43\% & 73.81\% \\
DeepSeek-V4-Flash & 44.54\% & 46.73\% & 66.67\% & 69.05\% \\
Mistral Medium & 51.31\% & 53.53\% & 61.90\% & 66.67\% \\
\hline
\end{tabular}
\end{center}
\end{table}

\begin{figure}[!t]
    \centering
    \includegraphics[width=0.83\linewidth]{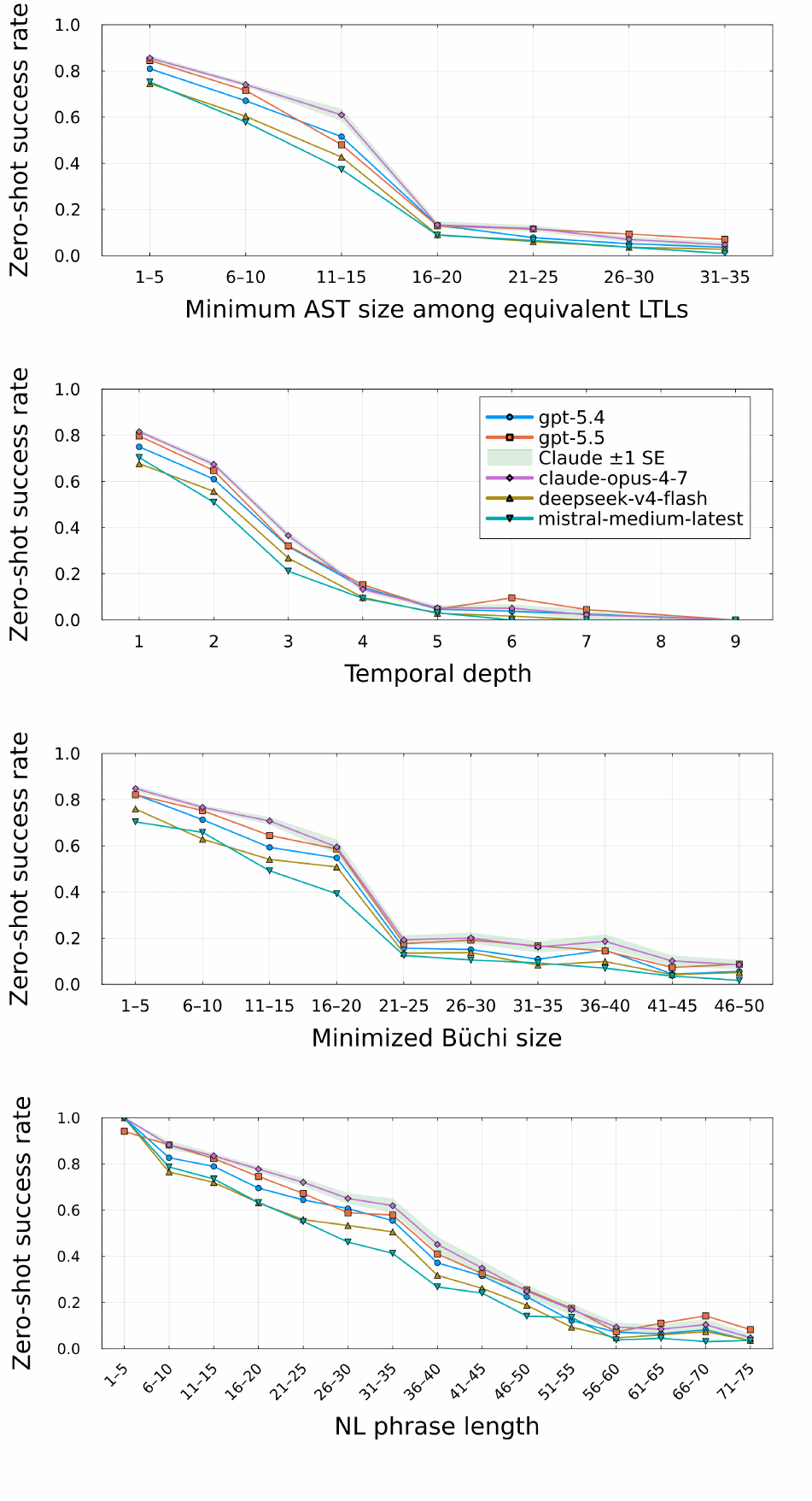}
    \caption{Zero-shot success rate of the evaluated LLMs, as defined in \eqref{eq:success_rate}, shown as a function of the minimum AST size among equivalent benchmark formulas, temporal depth, minimized B\"uchi automaton size, and NL input length.}
    \label{fig:success_rate_vs_size}
\end{figure}

Figure~\ref{fig:success_rate_vs_size} demonstrates that the zero-shot success rate
decreases as a function of four measures: minimum AST size among semantically
equivalent LTL formulas in the benchmark, temporal depth, minimized
nondeterministic B\"uchi automaton size~\cite{duret2022spot},
and NL input length. The first two characterize the structural complexity of
the target formula, while the B\"uchi automaton size provides an automata-based
proxy for the semantic complexity of the specification. NL input length captures the input's surface complexity. The minimum-AST
measure assigns to each formula the smallest AST size
among semantically equivalent formulas in the benchmark,
reducing dependence on a particular syntactic representation.
Few-shot curves are omitted because they follow similar trends.

To quantify the conditional associations of the complexity predictors with
translation success, we fit a multivariable logistic regression model for
each LLM, with translation success as the binary response and the
standardized predictors given by the four measures shown in
Fig.~\ref{fig:success_rate_vs_size}, together with raw AST size. The models
are fitted using iteratively reweighted least squares. Each coefficient
therefore describes the association of the corresponding predictor with
translation success while holding the remaining predictors fixed. The
fitted models achieve McFadden's $R^2$ values of 0.19--0.34 and areas under
the receiver operating characteristic curve of 0.77--0.86.

Because each formula contributes four correlated NL variants, or eight when
a PNF rewriting is also available, this dependence may lead to underestimated
standard errors. We therefore use cluster-robust standard errors, grouping paraphrases that correspond to semantically equivalent formulas into the same cluster.\footnote{CR1S correction; see the \texttt{CovarianceMatrices.jl} documentation.}
Table~\ref{tab:complexity_analysis} reports the combined logistic regression coefficients obtained through random-effects meta-analysis based on these cluster-robust standard errors, together with their 95\% confidence intervals, between-model heterogeneity, and the number of LLMs for which
each predictor is statistically significant. Overall, minimized B\"uchi automaton size, minimum AST size, and temporal depth show consistently negative and largely significant associations with translation success, whereas NL input length is less consistently significant across the evaluated LLMs.

\begin{table}[t]
\caption{Coefficients for the complexity measures,
combined through random-effects meta-analysis using cluster-robust standard errors. Reported are significance across the five LLMs, 95\% confidence intervals, between-model heterogeneity, and the range of coefficients across LLMs.}
\label{tab:complexity_analysis}
\centering
\scriptsize
\setlength{\tabcolsep}{1.2pt}

\begin{tabular}{|p{65pt}|p{16pt}|p{69pt}|p{20pt}|p{58pt}|}
\hline
\textbf{Measure} &
\textbf{{Sig.}} &
\textbf{{Coefficient [95\% CI]}} &
\textbf{{$I^2$}} &
\textbf{{Range}} \\
\hline
B\"uchi automaton size &
4/5 &
$-0.815$ $[-1.169,\,-0.460]$ &
$60.1\%$ &
$[-1.360,\,-0.319]$ \\
\hline
Minimum AST size &
5/5 &
$-0.438$ $[-0.512,\,-0.363]$ &
$0.0\%$ &
$[-0.504,\,-0.357]$ \\
\hline
Temporal depth &
5/5 &
$-0.257$ $[-0.336,\,-0.178]$ &
$49.0\%$ &
$[-0.362,\,-0.151]$ \\
\hline
NL input length &
1/5 &
$-0.090$ $[-0.168,\,-0.012]$ &
$26.1\%$ &
$[-0.249,\,0.012]$ \\
\hline
Raw AST size &
0/5 &
$-0.088$ $[-0.171,\,-0.005]$ &
$0.0\%$ &
$[-0.145,\,-0.020]$ \\
\hline
\end{tabular}
\end{table}

\subsection{NL Interface for ABCD}
\label{sec:interface}

\begin{figure}[!t]
    \centering
    \includegraphics[width=0.99\linewidth]{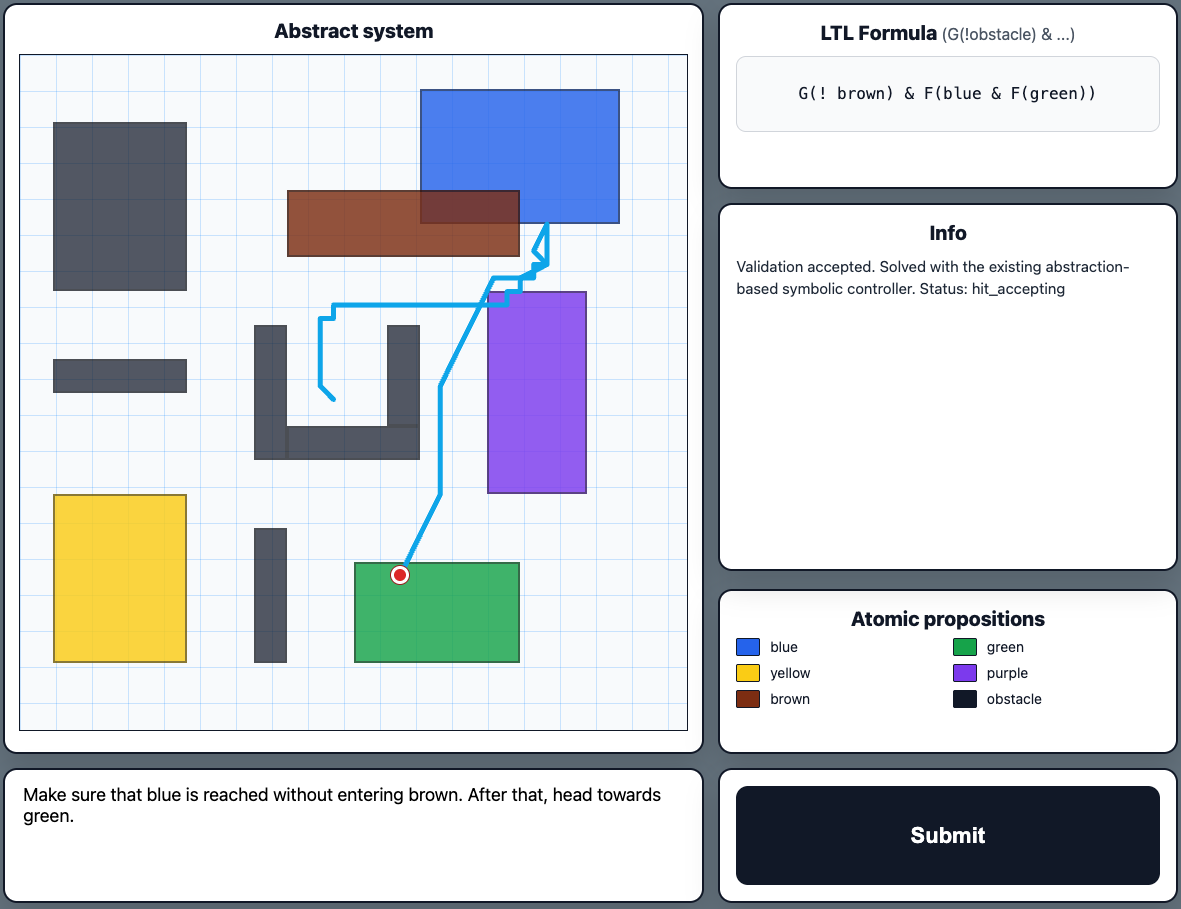}
    \caption{NL interface for ABCD.}
    \label{fig:interface}
\end{figure}

Figure~\ref{fig:interface} shows the developed interactive interface for
synthesizing ABCD controllers from NL requirements, demonstrated on the
two-wheeled vehicle model of~\cite{reissig2016feedback}. The interface
displays the symbolic abstraction and atomic propositions, accepts an NL
requirement, shows the generated LTL formula, and reports the pipeline
status.

To evaluate NL-to-LTL translation in this setting, we constructed
a task-specific dataset from the abstract benchmark. For each LTL template, the abstract atomic propositions and
corresponding NL descriptions were mapped to the propositions
of the example abstraction in two ways, yielding
two task-specific requirements per template. We retained only formulas whose logical constraints were
compatible with the labeled regions of the abstraction;
for instance, \(\Diamond(\textit{blue}\wedge\textit{green})\)
was discarded because
the \textit{blue} and \textit{green} regions are disjoint.
The resulting NL requirements were paraphrased using GPT-5.4-mini to improve
fluency while preserving their temporal meaning.
Claude Opus 4.7 achieved a success rate of 48.37\%. This lower rate is consistent with results reported in
VLTL~\cite{english2025verifiable}, as task-specific translation
additionally requires grounding propositions and interpreting
task-specific actions.

We further evaluated the correctly translated requirements supported by
the current ABCD synthesis backend. Among these requirements, a controller with a nonempty winning set was
synthesized for 97.27\%, with a 100\% specification-satisfaction rate
among the synthesized controllers.

\section{Conclusion}

This paper presented an LLM-enhanced pipeline for ABCD from NL requirements,
integrating NL-to-LTL translation, human-in-the-loop validation, and symbolic
synthesis in \texttt{Dionysos}. The primary quantitative
benchmark focuses on NL-to-LTL translation, while the complete
NL-to-controller pipeline is additionally evaluated on a task-specific
control example. We introduced a benchmark with broad logical coverage,
linguistic variation, and multiple measures of specification complexity.
Experiments with state-of-the-art LLMs showed that translation accuracy
decreases as target-specification complexity increases.
The downstream evaluation further demonstrates the
integration of translation with symbolic controller synthesis and formal
specification satisfaction. Future work will assess how
NL-to-LTL translation errors and human-in-the-loop validation influence
controller-synthesis outcomes.

\bibliographystyle{IEEEtran} 
\bibliography{refs}

\appendices

\section{Prompts}
\label{app:paraphrasing_prompt}

\subsection{Paraphrasing Prompt}
The verbatim statements are paraphrased using the prompt below with three
different LLMs: GPT-5.4-mini, Gemini-2.5-Flash, and DeepSeek-Chat. For
this stage, we set the LLM sampling temperature to $T=0.6$ to encourage
more diverse paraphrases while keeping the temperature sufficiently low to
reduce hallucinations and preserve semantic equivalence~\cite{li2025exploring}.

\begin{promptbox}
\setlength{\parindent}{0pt}
\small

You are given a natural-language statement derived from an LTL formula.

Your task is to produce \textbf{one} paraphrase that:
\begin{itemize}
    \item preserves the exact logical meaning,
    \item remains faithful to the original statement,
    \item sounds natural in English,
    \item uses noticeably different wording and, when possible, a different sentence structure.
\end{itemize}

\textbf{Strict rules:}
\begin{enumerate}
    \item Do not change the meaning.
    \item Do not add or remove conditions.
    \item Do not introduce domain-specific assumptions.
    \item Keep proposition names unchanged, e.g., \texttt{prop\_1}, \texttt{prop\_2}.
    \item Prefer genuine rewording over minor edits.
    \item Avoid copying the original sentence structure unless necessary.
    \item Output only the paraphrased sentence.
\end{enumerate}

\textbf{Input statement:}

\texttt{\$INPUT\_SENTENCE\$}

\textbf{Paraphrase:}
\end{promptbox}

\subsection{NL-to-LTL Prompt}
\label{sec:nltl_prompt}

This section presents the prompt used for NL-to-LTL translation. To ensure
reproducibility and because the task requires deterministic and
semantically faithful outputs, we set the sampling temperature to $T=0$
for the evaluated LLMs that expose a temperature parameter. For zero-shot
evaluation, we use the same prompt without examples. For few-shot
evaluation, we include the representative NL/LTL pairs provided below
before presenting the target input.

\vspace{0.5em}
\begin{promptbox}
\small
\setlength{\parindent}{0pt}

You translate natural-language statements into Linear Temporal Logic (LTL).

\textbf{Output rules:}
\begin{itemize}
    \item Output only the LTL formula.
    \item Do not add explanations, comments, labels, or code fences.
    \item Keep proposition names exactly as they appear (e.g., \texttt{prop\_1}, \texttt{prop\_2}).
    \item Preserve the exact logical and temporal meaning.
    \item Do not introduce domain assumptions.
    \item Do not use any proposition outside the available proposition names in the input.
    \item Use only the operators \texttt{!} (not), \texttt{\&} (and), \texttt{|} (or), \texttt{->} (implication), \texttt{<->} (equivalence), \texttt{X} (next), \texttt{F} (Finally), \texttt{G} (Globally), and \texttt{U} (Until).
    \item Use parentheses to avoid ambiguity.
\end{itemize}

\vspace{0.5em}
\textbf{Examples for the few-shot setting:}

\vspace{0.3em}
\textbf{Natural-language statement:}

Globally, if \texttt{prop\_1} holds, then \texttt{prop\_3} is true until \texttt{prop\_2}.

\textbf{LTL:}

\texttt{G (prop\_1 -> (prop\_3 U prop\_2))}

\vspace{0.4em}
\textbf{Natural-language statement:}

If \texttt{prop\_1} holds, then from that point on \texttt{prop\_2} must always hold.

\textbf{LTL:}

\texttt{(prop\_1 -> G (prop\_2))}

\vspace{0.4em}
\textbf{Natural-language statement:}

\texttt{prop\_1} holds exactly when \texttt{prop\_2} always holds.

\textbf{LTL:}

\texttt{(prop\_1 <-> G (prop\_2))}

\vspace{0.4em}
\textbf{Natural-language statement:}

Eventually \texttt{prop\_1} holds, and after that eventually \texttt{prop\_2} holds.

\textbf{LTL:}

\texttt{F (prop\_1 \& F (prop\_2))}

\vspace{0.4em}
\textbf{Natural-language statement:}

\texttt{prop\_1} must never occur.

\textbf{LTL:}

\texttt{G (!prop\_1)}

\vspace{0.4em}
\textbf{Natural-language statement:}

\texttt{prop\_1} holds until \texttt{prop\_2} occurs.

\textbf{LTL:}

\texttt{(prop\_1 U prop\_2)}

\vspace{0.5em}
\textbf{Target input:}

\textbf{Natural-language statement:}

\texttt{\$INPUT\_SENTENCE\$}

\textbf{LTL:}
\end{promptbox}
\vspace{0.5em}

\subsection{Prompt for Task-Specific Instruction Rewriting}
\label{sec:task_specific_rewriting_prompt}

This section presents the prompt used to rewrite NL paraphrases after
instantiating the abstract atomic propositions with task-specific
propositions from the robot navigation environment.

\vspace{0.5em}
\begin{promptbox}
\small
\setlength{\parindent}{0pt}

You are given a natural-language description derived from an LTL formula
for a robot navigation task.

Rewrite the description as a natural task instruction that a human might
give to a robot.

The environment consists of colored regions. A proposition such as
``blue is true'' means that the robot is in the blue region. Likewise for
brown, yellow, green, and purple.

\vspace{0.5em}
\textbf{Rules:}
\begin{itemize}
    \item Preserve the exact meaning of the original sentence.
    \item Do not introduce new goals, constraints, or assumptions.
    \item Do not mention LTL, propositions, truth values, formulas, or logic.
    \item Use natural navigation language.
    \item Keep the instruction concise, one sentence whenever possible.
    \item Keep some of the input sentence's phrasing style when possible, while still rewriting it as a natural robot instruction.
    \item Preserve stylistic diversity across different inputs: if two inputs are phrased differently but mean the same thing, do not automatically rewrite them using the exact same wording.
    \item Avoid always defaulting to the same verb pattern such as only ``reach'' or only ``visit'' when another equally correct natural phrasing is possible.
    \item Keep the rewriting natural, but retain differences in tone or structure when the input wording differs.
    \item Do not explain your reasoning.
    \item Output only the rewritten instruction.
\end{itemize}

\textbf{Examples:}

\textbf{Example 1}\\
Input: \texttt{"blue eventually becomes true."}\\
Output: \texttt{"Eventually reach the blue region."}

\vspace{0.3em}
\textbf{Example 2}\\
Input: \texttt{"brown becomes true infinitely often."}\\
Output: \texttt{"Visit the brown region infinitely often."}

\vspace{0.3em}
\textbf{Example 3}\\
Input: \texttt{"If green is true, then yellow must always be true."}\\
Output: \texttt{"If the robot is in the green region, then it must always stay in the yellow region."}

\vspace{0.3em}
\textbf{Example 4}\\
Input: \texttt{"Either brown is true, or purple eventually becomes true."}\\
Output: \texttt{"Either the robot is in the brown region, or it should eventually reach the purple region."}

\vspace{0.3em}
\textbf{Example 5}\\
Input: \texttt{"In the next step, green is true."}\\
Output: \texttt{"Move so that the robot is in the green region at the next step."}

\vspace{0.5em}
\textbf{Rewrite the following sentence:}

\texttt{\{natural\_paraphrase\}}

\end{promptbox}
\vspace{0.5em}

\end{document}